\documentclass[summary]{gass2017}

\title{A Highly-Sensitive Cryogenic Phased Array Feed for the Green Bank Telescope}

\author{D. Anish Roshi\affref{ref1}, 
W. Shillue\affref{ref1}, 
J. R. Fisher\affref{ref1}, 
M. Morgan\affref{ref1}, 
J. Castro\affref{ref1}, 
W. Groves\affref{ref1},
T. Boyd\affref{ref1},
B. Simon\affref{ref2}, \\
L. Hawkins\affref{ref2},
V. van Tonder\affref{ref2}, 
J. D. Nelson\affref{ref2},
J. Ray\affref{ref2}, 
T. Chamberlain\affref{ref2},
S. White\affref{ref2},
R. Black\affref{ref3}, 
K. F. Warnick\affref{ref3}, \\
B. Jeffs\affref{ref3},
and R. Prestage\affref{ref2}}

\affiliation{%
\aff{ref1}{National Radio Astronomy Observatory (NRAO), 520 Edgemont Road, 
Charlottesville, VA 22903, USA.
}
\aff{ref2}{Green Bank Observatory (GBO), 155 Observatory Rd,
Green Bank, WV 24944, USA.}
\aff{ref3}{ECE Dept., Brigham Young University (BYU), Provo, UT 84602, USA.}
}

\begin{document}

\maketitle

\begin{abstract}
In this paper, we describe the development of a new L-band (1.4 GHz) Cryogenic Phased 
Array Feed (PAF) system, referred to as the GBT2 array. Results from initial measurements 
made with the GBT2 array are also presented. The PAF was developed for the Green Bank
Telescope (GBT) as part of the Focal L-band Array for the GBT (FLAG) project.
During the first stage of the
development work (Phase I), a prototype cryogenic 19 element dual-polarized array with 
``Kite'' dipole elements  was 
developed and tested on the GBT. The measured system temperature over efficiency 
($T_{sys}/\eta$) ratio for the bore sight beam of the Kite array was 45.5 K at 1.55 GHz. 
The off-boresight $T_{sys}/\eta$ shows an increase by 13 K at an offset equal to 
the half power beam width  (7$^{'}$.2 at 1.7 GHz). Our 
measurements indicate that the off-boresight degradation and field-of-view (FoV) limitation of the Kite array is simply due to the fixed array size.  To increase the FoV, a new 19-element GBT2 array with larger 
array spacing was developed during FLAG Phase II. The frequency response of the array was optimized 
from 1.2 to 1.6 GHz. A system with larger cryostat, new low noise amplifiers (LNAs),
down-conversion and digitization close to the front end, unformatted digital 
transmission over fiber, ROACH II based polyphase filter banks (PFBs) with bandwidth 150 MHz
and a data acquisition system that records voltage samples from one of the PFB channels
were all developed. The data presented here is processed off-line. The receiver temperature measured (off the telescope, on cold sky/hot load, with no beamforming) with 
the new system is 17 K at 1.4 GHz, an improvement $>$ 8 K over the previous Kite array. 
Measurements with the GBT2 array on the telescope are in progress.  
A real time 150 MHz beamformer is also being developed as part of an NSF-funded 
collaboration between NRAO/GBO/BYU \& West Virginia University (Beamformer Project)
to support science observations.
\end{abstract}

\section{Introduction}

For the advancement of Radio Astronomy,  a large class of observations 
requires extended and large scale surveys.  However, large
reflector antennas used by radio astronomers have small field of view (FoV) limiting
the survey speed. Multiple optimized feeds conventionally used to increase
the FoV, however, do not produce overlapping beams and the off-axis feeds 
suffer efficiency degradation, both resulting in reduced mapping efficiency. 
The Phased array feed is another technology that can be used to increase the 
FoV\cite{rfisher1996}. PAFs consist of densely packed, electrically 
small feed arrays, each of which  over illuminates the reflector. Signals from the 
array elements are combined to form multiple, overlapping beams with low spillover noise. 
A low-noise, multi-beam PAF system increases the survey speed of the telescope, 
enabling radio astronomers to make sensitive, large scale surveys. 

Developing a low-noise PAF is difficult due to mutual coupling between array elements. 
Despite this difficulty several groups across the world
have undertaken development of low-noise PAFs (see \cite{karl2016}). 
The development work is being done for both
single radio telescopes and interferometers. The APERture Tile In Focus (APERTIF)\cite{oosterlooetal2010} and
Chequerboard PAFs are un-cooled, broadband ($>$ 800 MHz) PAFs near 1.4 GHz installed
respectively at the Westerbork Synthesis Radio Telescope (WSRT) and Australian Square
Kilometer Prototype (ASKAP)\cite{chippendale2015}. The Dominion Radio Astrophysical Observatory (DRAO) used
Vivaldi elements similar to APERTIF for a PHased Array feed Demonstrator (PHAD)\cite{gray2011}.
PAFs that are being developed for single dish telescopes are cooled arrays. They include AO40 
being developed for the Arecibo telescope by Cornell University\cite{cortes2015} 
and a PAF for the Five hundred meter
Aperture Spherical Telescope (FAST)\cite{Wu2016}, both operating near 1.4 GHz. A higher frequency (70 -- 95 GHz)
PAF is being developed by the University of Massachusetts for the GBT\cite{erickson2015}.   

The Focal L-band Array for the GBT (FLAG) is a collaborative project between NRAO, GBO 
\& BYU. This collaboration has developed a detailed model for the PAF system and
the associated beamforming algorithms\cite{warnicketal2009, jeffs2008}. 
The PAF modeling efforts of BYU led to a design procedure 
that will iteratively reshape the dipole structure in order to optimize the 
sensitivity of the array feed on a reflector telescope\cite{warnicketal2011}. A prototype
array, referred to as the Kite array, was constructed and tested on the GBT, which
forms Phase I of the FLAG project. Section~\ref{ph1} summarizes the Phase I developments and results. 
During this phase NRAO also developed an independent PAF model, which is described in 
Section~\ref{phm}. The collaboration then developed the GBT2 dipole array,
which forms FLAG Phase II. The dipole spacing in GBT2 array
is optimized to obtain 7 beams on the GBT spaced at the half power beam width over its operating
frequency. The full system has been redesigned as part of Phase II work.
Section~\ref{ph2} gives a description of the system and the initial measurement results. 

\section{FLAG Phase I: Development, Measurement and Results}
\label{ph1}

\begin{figure}[htbp]
  \centering
\begin{tabular}{cc}
  \includegraphics[width=37mm]{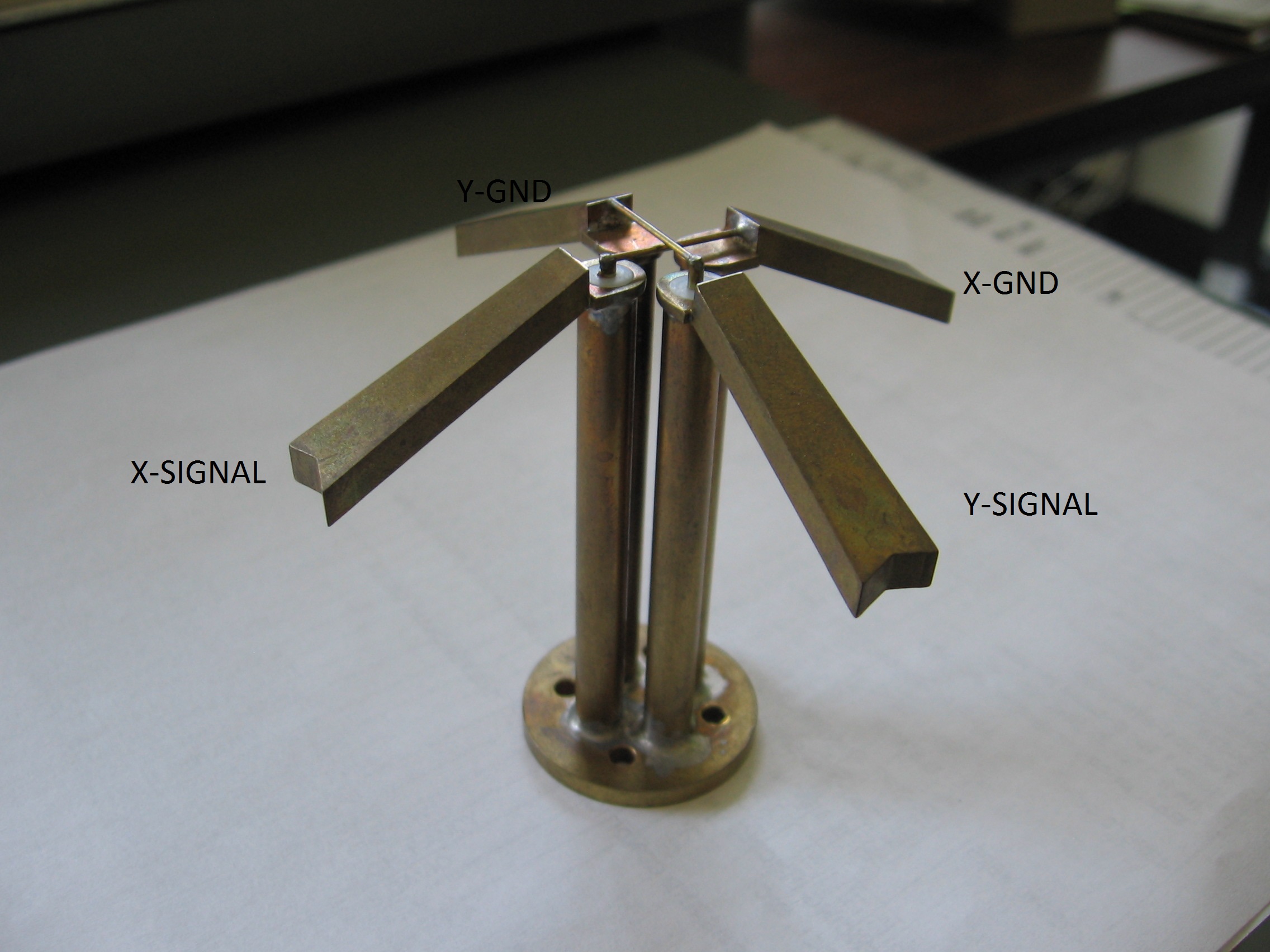} &
  \includegraphics[height=28mm, width=37mm]{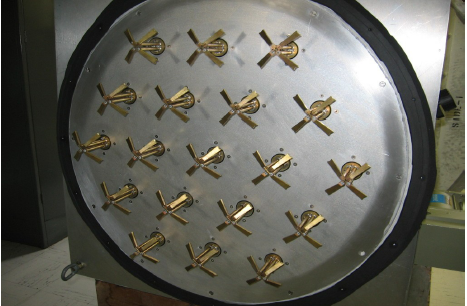} 
\end{tabular}
  \caption{{\bf Left :} The Kite dipole. {\bf Right:} The 19 element dual polarized Kite array
developed during Phase I. The spacing between the dipoles is 12 cm (0.56$\lambda$ at 1.4 GHz). 
    }
  \label{fig1}
\end{figure}

\begin{figure}[htbp]
  \centering
\begin{tabular}{cc}
  \includegraphics[height=32mm,width=37mm]{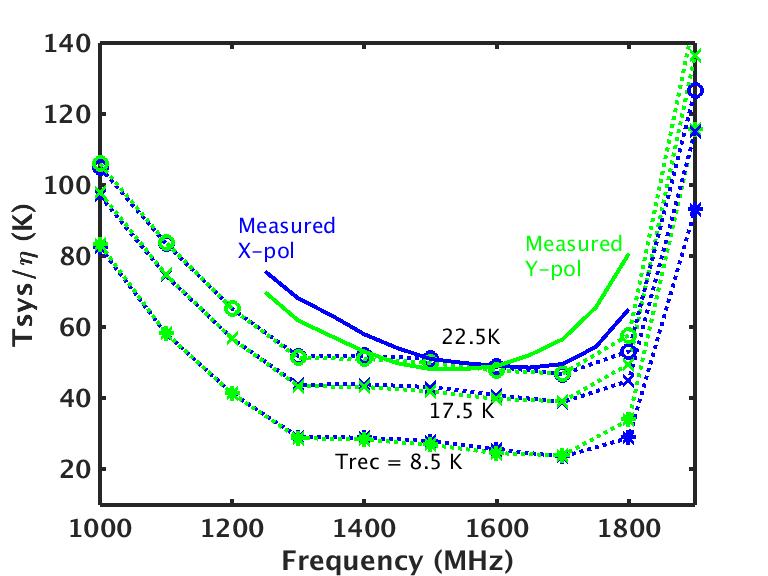} &
  \includegraphics[height=32mm, width=37mm]{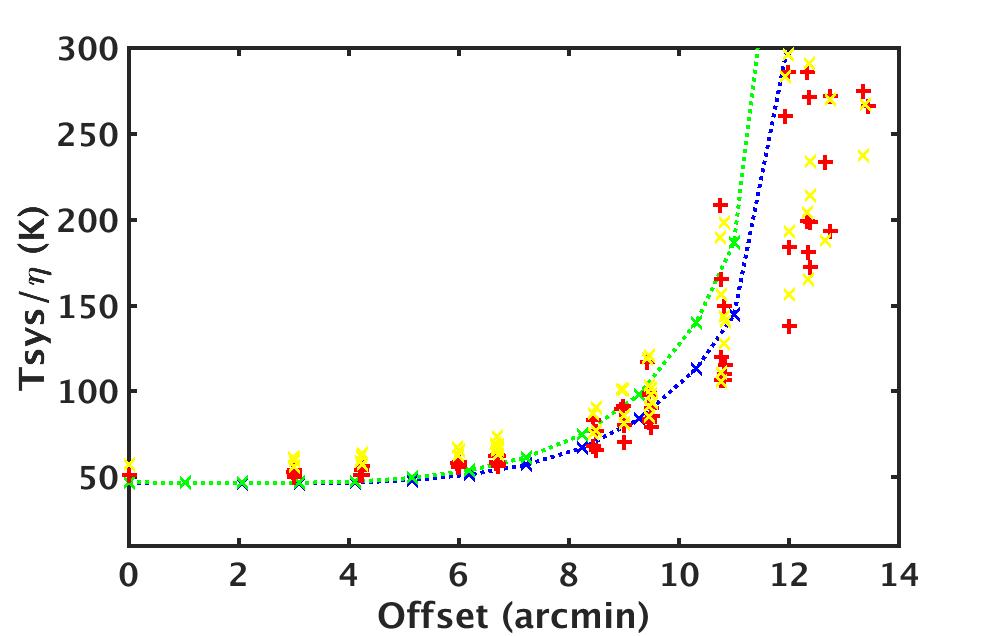}  
\end{tabular}
\caption{{\bf Left :} Measured bore sight $T_{sys}/\eta$ vs frequency for the Kite array on the GBT 
(solid lines) along with the NRAO PAF model predictions (dotted curves). Beams for these measurements
are formed by maximizing SNR on the source Virgo A. The X and Y polarization are shown
in blue and green respectively. The model predictions are presented for the receiver temperatures 
due to the LNA alone (8.5 K at 1.5 GHz) and with an increased value to account for extra noise
contributions in the system (see text).  
{\bf Right:} $T_{sys}/\eta$ vs offset from bore sight obtained from Virgo A observations at 1.7 GHz 
are indicated by `red +' (X polarization) and `yellow x' (Y polarization). 
Model predictions at 1.7 GHz are indicated by the dotted curve(blue -- X polarization, green -- Y polarization). 
}
\label{fig2}
\end{figure}

\begin{figure}[htbp]
  \centering
\begin{tabular}{cc}
  \includegraphics[width=37mm]{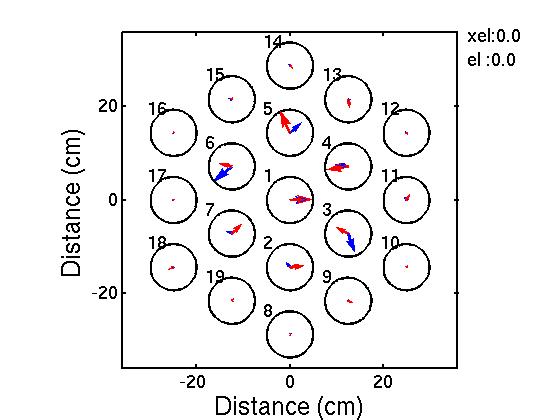} &
  \includegraphics[width=37mm]{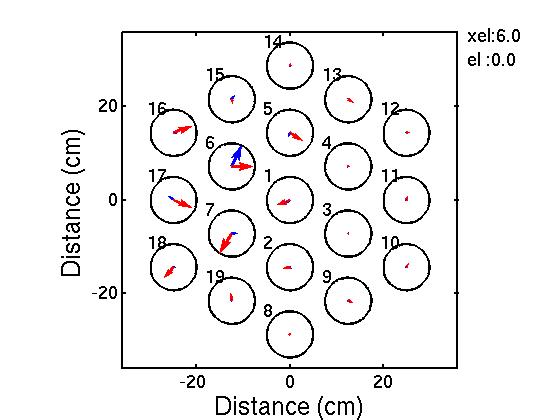} 
\end{tabular}
  \caption{
{\bf Left:} The maximum SNR weights for the bore sight beam are shown on the PAF
layout. The weights are shown on a unit circle with length of the arrow representing the
magnitude of the complex value and the angle from the horizontal representing its phase
(blue -- X-polarization, red -- Y-polarization).  As seen in the
figure significant values for the weights are applied to the 7 central elements to form the 
bore sight beam 
{\bf Right:} The weights for the beam at 6$^{'}$ offset in cross-elevation from the bore sight.               
The elements with significant values for the weights are at the edge of the array, indicating
that the FoV of the PAF is limited by the array size.
}
\label{fig3}
\end{figure}

The Kite array developed during FLAG Phase I is shown in Fig.~\ref{fig1}. 
The dipoles were optimized for active impedance match to the LNA and for
maximum sensitivity on a reflector antenna over a specified FoV. 
The optimization was done over the frequency range 1.3 to 1.7 GHz. 
The design and fabrication of the dipoles were done by Karl Warnick 
and a team of students at BYU. The dipoles and balun, which 
were maintained at ambient temperature, were connected
to the LNA through a low-loss but high thermal impedance coaxial cable. 
The LNAs were located
inside the cryostat and cooled to 15K. The cryostat was developed by Roger 
Norrod at NRAO\cite{norrod2010}. Signal received by each dipole was 
amplified and transmitted through analog fiber optic links to an analog 
down-converter. The bandwidth at the output of the down-converter was 
reduced to about 400 KHz. The signals were then digitized using 12-bit 
ADCs and the voltage samples were recorded to disk. All processing
was done offline on the recorded data, where we formed beams by 
maximizing the signal-to-noise ratio (SNR)\cite{roshi2015}.

In early 2015, the receiver temperature of the Kite array was measured at the
outdoor facility at GBO. The receiver temperature (median
value of the temperatures of the 38 dipoles) as a function of frequency 
is shown in Fig.~\ref{fig6}. The best value obtained was 25 K at 1.7 GHz. 
The array was then installed on the GBT and an extensive
set of measurements were made\cite{roshi2015}. We express the performance of the PAF in terms
of the ratio $T_{sys}/\eta$, where $T_{sys}$ is the system temperature and
$\eta$ is approximately the aperture efficiency if the radiation efficiency is 
close to unity. Fig.~\ref{fig2} shows an example measurement result made using the
source Virgo A. The best value obtained for the boresight beam $T_{sys}/\eta$ 
is 45.5 K near 1.55 GHz. The FoV of the array is measured at 1.7 GHz by observing Virgo A
at different offsets from the boresight. As seen in Fig.~\ref{fig2}, the 
$T_{sys}/\eta$  increases by about 13 K at an offset equal to the half power beam width 
(7$^{'}$.2 at 1.7 GHz). 
Our measurements show that the FoV is limited by the finite size of the array. 
This is illustrated in Fig.~\ref{fig3}, where we plot the maximum SNR weights for 
boresight and those for an offset of 6$^{'}$. The 7 elements which have significant
values for the weights are located at the edge of the array for 6$^{'}$ offset.
Thus, for this array, an additional ring of elements would be needed for offsets larger than $\sim$ 6$^{'}$
in order to form a high sensitivity beam and therefore the FoV is limited by the array size.

\section{The NRAO PAF model}
\label{phm}

A PAF model was independently developed at NRAO. Our modeling starts with the
electromagnetic simulation of the full array in the Computer Simulation Technology (CST) 
software package. This simulation provides the S-parameters and field patterns of each dipole when embedded 
in the array. The outputs of the CST are then used to compute the embedded beam patterns, 
which are defined as the field pattern when a dipole in the array is excited and 
all other dipole inputs are short circuited. The embedded beam patterns and
the S-parameters along with the LNA noise model are used to predict  
$T_{sys}/\eta$ when the PAF is place on the GBT. Any additional noise 
contribution (like losses in the system) is accounted in the model 
by adding a noise temperature to the receiver temperature due to LNA alone.
A complete description of the model starting from the first principles 
is available as an NRAO internal report\cite{roshi2016}. In Fig.~\ref{fig2},
we compare the model predictions with measurements made on the GBT. As seen in
Fig.~\ref{fig2}, the model well predicts the off-boresight measurements
and to a lesser extent the frequency dependence of the boresight measurements.  
An excess noise temperature of 14 K at 1.5 GHz over the LNA noise
contribution (noise temperature due to LNA is 8.5 K) is required 
for the model to match with the measurements. 
Out of this excess, the total contribution from losses in the system and the expected additional
noise in the LNA is approximately 9 K. The origin of the unaccountable 5 K noise contribution 
needs further investigation.

\section{FLAG Phase II: Development, Measurement and Results}
\label{ph2}

\begin{figure}[htbp]
  \centering
  \includegraphics[height=30mm]{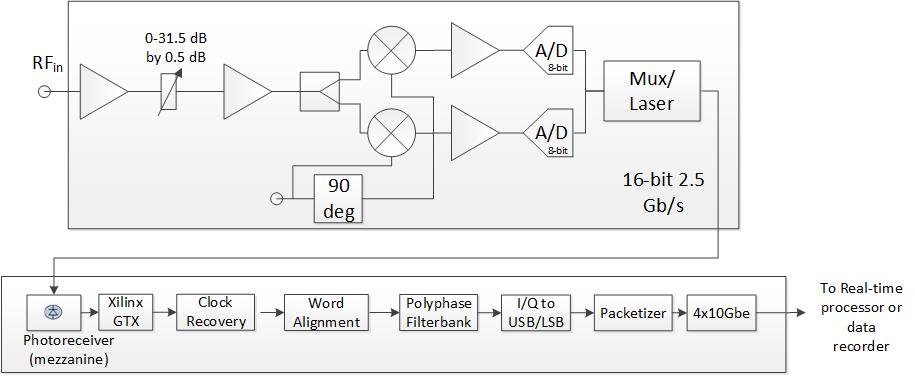}  
  \caption{Block diagram of the GBT2 system developed during the FLAG Phase II.
    }
  \label{fig5}
\end{figure}

\begin{figure}[htbp]
  \centering
\begin{tabular}{cc}
  \includegraphics[height=33mm,width=37mm]{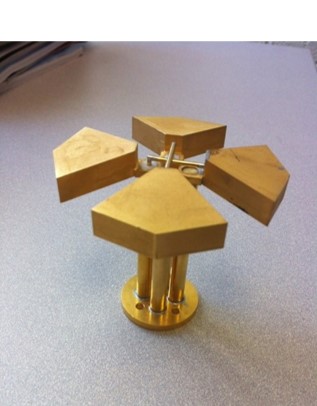} & 
  \includegraphics[height=29mm, width=37mm]{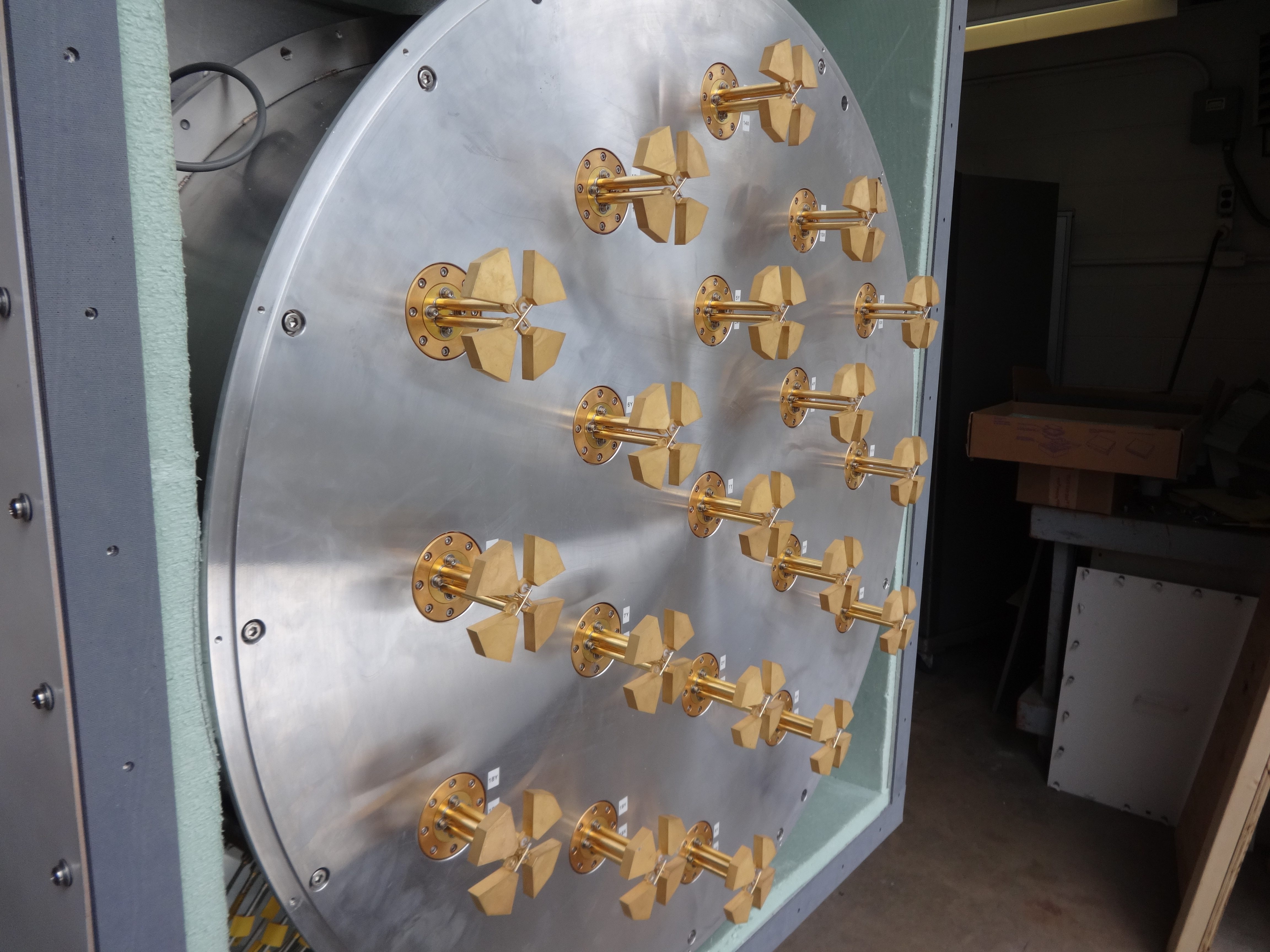} \\ 
  \includegraphics[width=37mm]{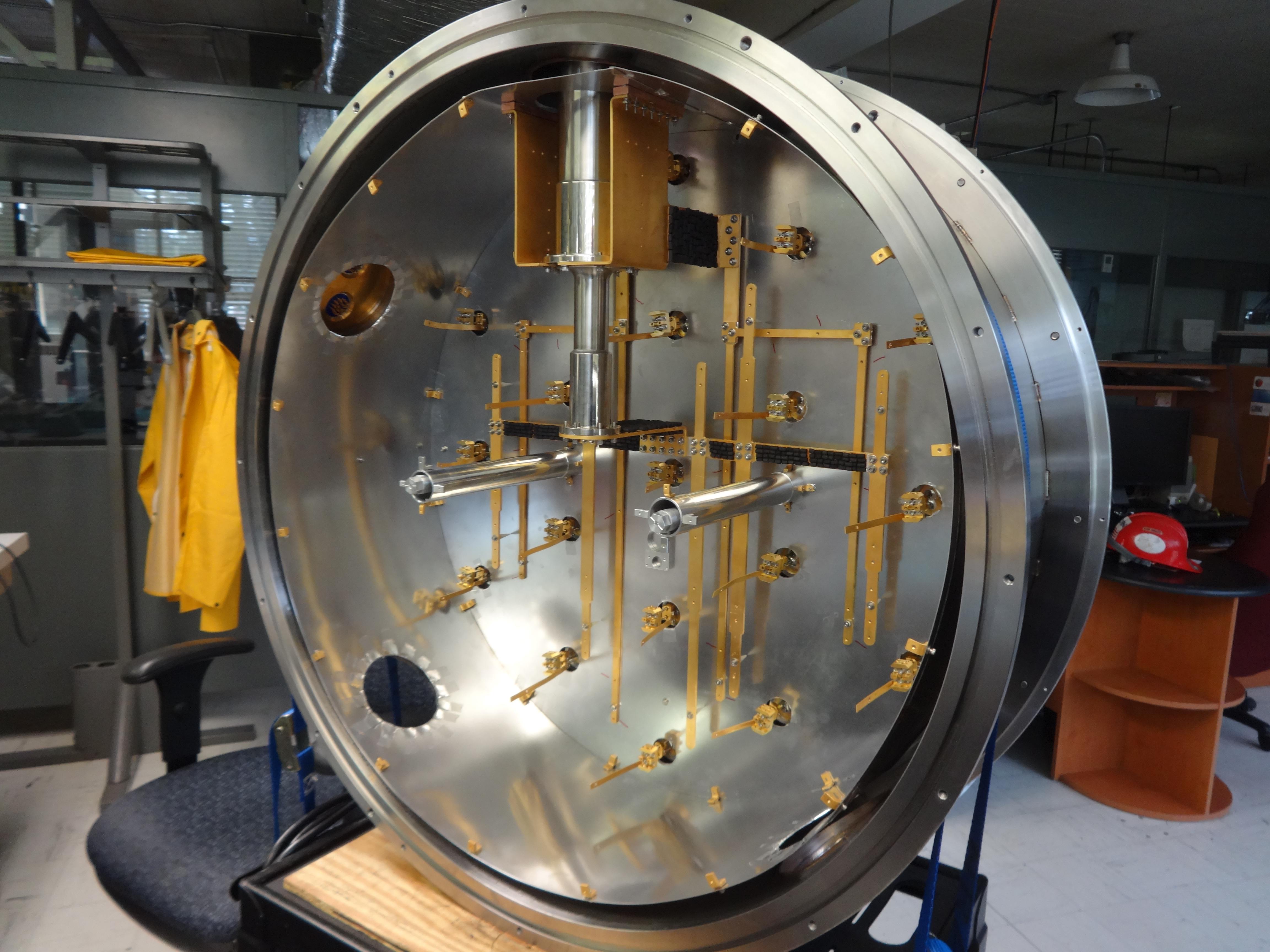} &
  \includegraphics[width=37mm]{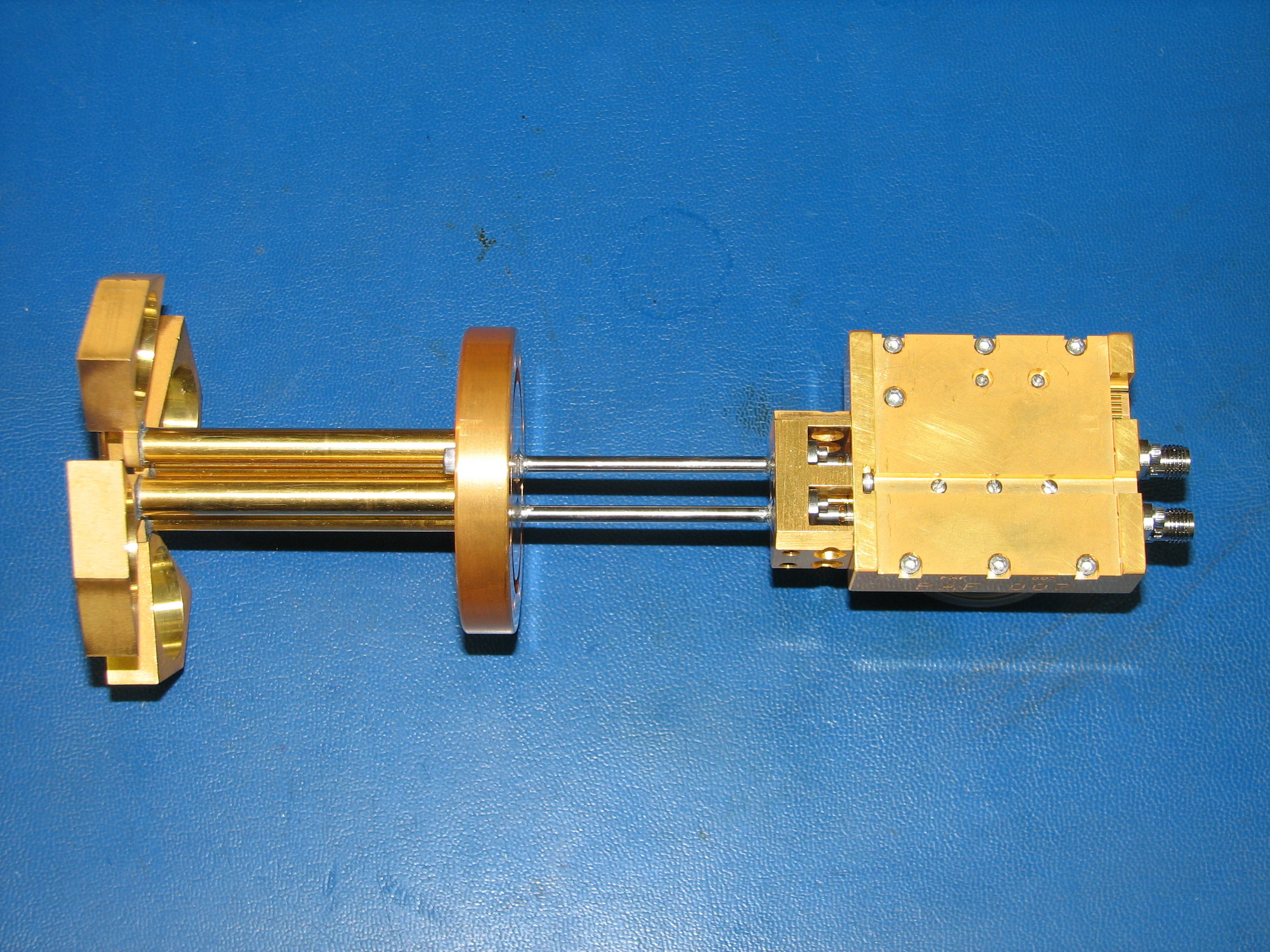} 
\end{tabular}
  \caption{{\bf Top left :} The GBT2 dipole. {\bf Top right:} The 19 element dual polarized GBT2 dipole array  
developed during FLAG Phase II. The spacing between the array elements is 15.6 cm (0.73$\lambda$ at 1.4 GHz). 
{\bf Bottom left:} Cryostat for the GBT2 array. {\bf Bottom right:} The new low-noise amplifier, 
the thermal transition and dipole assembly. 
}
  \label{fig4}
\end{figure}

\begin{figure}[htbp]
  \centering
\begin{tabular}{cc}
  \includegraphics[width=37mm]{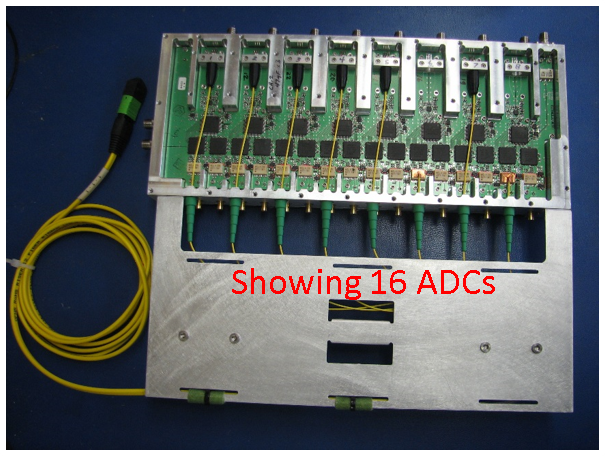} &
  \includegraphics[height=31mm,width=37mm]{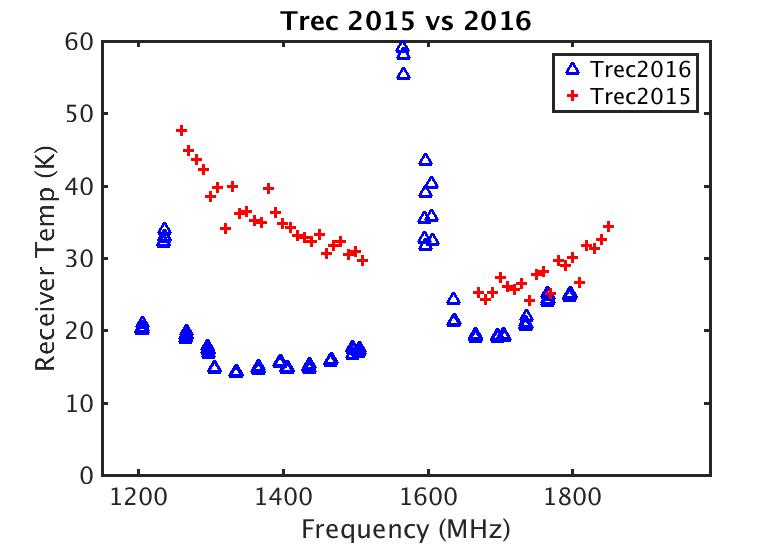} \\ 
\end{tabular}
  \caption{{\bf Left:} A subsystem, referred to as a `blade', of the integrated down-converter, 
digitizer and fiber optical link. Each `blade' can process 8 dipole outputs. 
{\bf Right:} Measured receiver temperature for the GBT2 array (blue) and Kite array (red). The measurements near 1.6 GHz are affected by radio frequency interference. 
    }
  \label{fig6}
\end{figure}

\begin{figure}[htbp]
  \centering
\begin{tabular}{cc}
  \includegraphics[height=32mm,width=37mm]{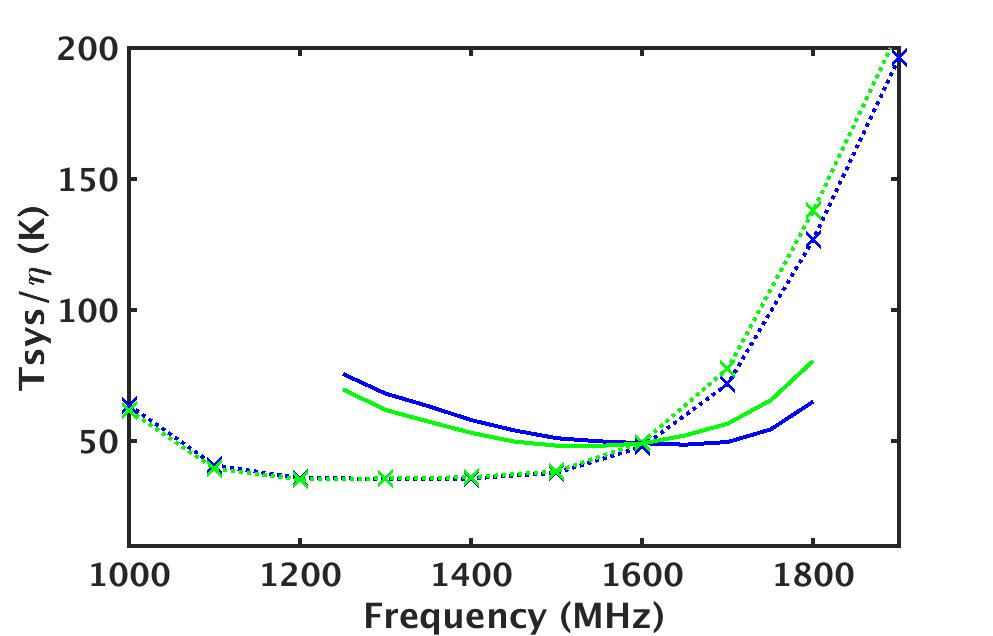} &
  \includegraphics[height=32mm,width=37mm]{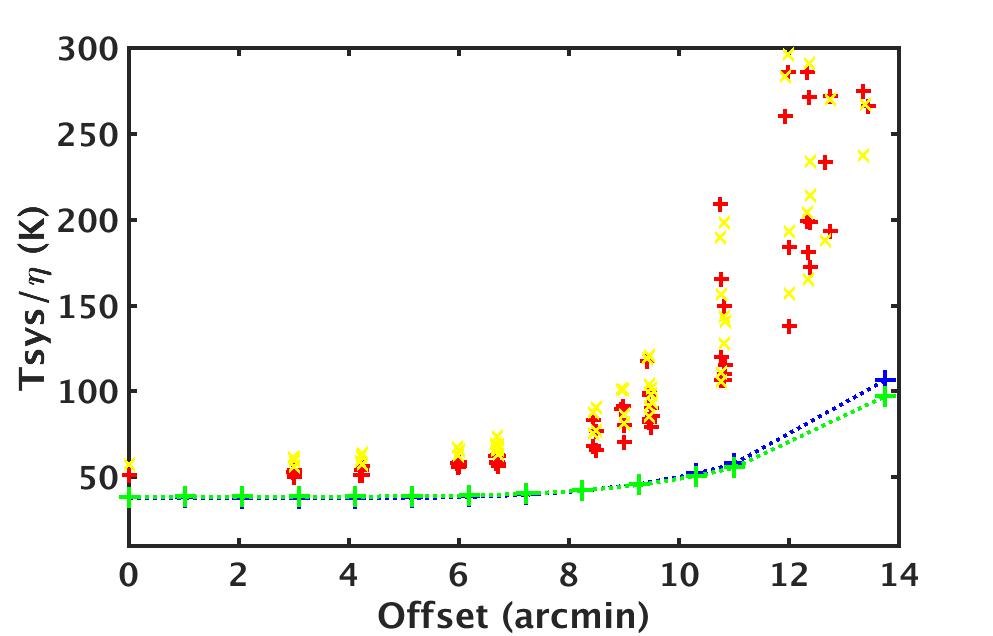} \\
\end{tabular}
  \caption{NRAO model predictions for the GBT2 array on the GBT. {\bf Left:} 
The expected $T_{sys}/\eta$ vs frequency for the bore sight beam (receiver temperature = 17 K at 1.5 GHz). 
The X and Y polarizations are shown in blue and green respectively. The measured performance 
of the Kite array is shown in solid line for comparison. {\bf Right:}  
The expected $T_{sys}/\eta$ vs offset from bore sight (dotted lines; blue -- X polarization, green
-- Y polarization) compared with the measured results from Kite array (red `+' \& yellow `x').
    }
  \label{fig7}
\end{figure}

Based on the success of the Phase I effort, the complete system was redesigned.
The block diagram of the new system is shown in Fig.~\ref{fig5}. To increase
the FoV of the PAF for the GBT, the GBT2 array (see Fig.~\ref{fig4}) was developed with 
an element spacing of 15.6 cm (0.73$\lambda$ at 1.4 GHz). 
The dipoles were redesigned by Karl Warnick for this  element spacing and the frequency
range was optimized to 1.2 to 1.6 GHz (see Fig.~\ref{fig7}). A larger cryostat was developed 
for the GBT2 array (see Fig.~\ref{fig4}) and the LNAs were redesigned for improved noise 
temperature\cite{mmorgan2016}. 
The signals are down converted and digitized with 8 bit ADCs
close to the front-end. The bandwidth at this stage is limited to 150 MHz. An unformatted 
digital link over optical fiber transports the samples
to the FPGA in ROACH II boards. Fig.~\ref{fig6} shows the digital down-converter 
and the link\cite{mmorgan2013}.
A polyphase filter bank (PFB) implemented in the FPGA breaks down the 150 MHz bandwidth
into 512 channels. The data acquisition system developed for system testing and commissioning 
of the front-end records voltage samples from one channel of the PFB to disk. The recorded
data is processed offline to form maximum SNR beams.

We have measured the receiver temperature of the system at the outdoor test 
facility at GBO, which is shown in Fig.~\ref{fig6}. The receiver temperature of GBT2 
array shows improvement by more than 8 K compared to that of
the Kite array (see Fig.~\ref{fig6}). The expected bore sight and 
off-bore sight performances of GBT2 array on the GBT obtained 
using the NRAO PAF model are shown in Fig.~\ref{fig7}. 
Measurements with the GBT2 array on the telescope is underway.
The Beamformer project will also be commissioning the real time system later this
year, which then will be used for regular science observations with the GBT. 

\section{Acknowledgments}

We acknowledge the help and support of L. Jensen, R. Dickman, S. Pan, W. Randolph, 
and K. O'Neil during the FLAG development work, testing and observations
on the GBT. The NRAO is a facility of the 
National Science Foundation (NSF) operated under cooperative agreement by Associated Universities, Inc.
We acknowledge the support of the NSF/ATI program through the awards 
1309832 and 1309815, 2013.

\end{document}